\documentclass[11pt]{article}

\usepackage[a4paper,margin=1in]{geometry}
\usepackage{authblk}
\usepackage{graphicx}
\usepackage{booktabs}
\usepackage{array}
\usepackage{longtable}
\usepackage{hyperref}
\usepackage{url}
\usepackage{amsmath}
\usepackage{enumitem}
\usepackage{xcolor}

\hypersetup{colorlinks=true, linkcolor=blue, citecolor=blue, urlcolor=blue}

\title{\textbf{LLMography: Transforming Human--AI Conversations into Traceability, Oversight, and Auditability Indicators}}

\author[1]{Mohammed Bousmah}
\affil[1]{LTI Laboratory, National School of Applied Sciences\\
Chouaib Doukkali University\\
El Jadida, Morocco\\
Email: \texttt{bousmah.m@ucd.ac.ma}}

\begin{document}

\maketitle

\begin{abstract}
The growing use of Large Language Models (LLMs) in education, software engineering, academic writing, and technical documentation raises a key question: how can we evaluate not only AI-assisted outputs, but also the interaction process that produced them? Current debates often focus on detecting whether a final artifact was generated by AI, while overlooking the conversation history that reveals human direction, AI contribution, corrections, validation, and traceability.

This paper introduces \textbf{LLMography}, a conceptual, terminological and computational framework for transforming Human--AI conversations into measurable indicators of provenance, human contribution, AI dependency, reproducibility, and auditability. By analogy with bibliography and webography, LLMography documents the dynamic trajectory of interaction between a human and a Large Language Model, not merely as a list of prompts, but as a structured trace of Human--AI co-production.

We present a prototype that analyzes Human--AI conversation traces and generates KPI reports including Prompt Quality Score, Human Direction Score, AI Dependency Level, Auditability Score, Final Output Traceability, Privacy Risk Level, and a recommended LLMography label. A preliminary exploratory evaluation was conducted on 19 anonymized audit reports from engineering students. The results show that most interactions were classified as \textit{Human--AI co-produced}, with average scores of 86.8/100 for Human Direction, 81.9/100 for Prompt Quality, 72.8/100 for Auditability, and 77.1/100 for Final Output Traceability.

The paper also applies LLMography to its own writing process. This self-application classified the manuscript preparation as \textbf{human-originated, human-directed, AI-assisted}, with strong human direction, moderate AI dependency, high auditability, and high traceability. The findings suggest that AI transparency should move beyond output detection toward documenting the history of interaction. In the AI age, outputs are not enough; we need the history of interaction.
\end{abstract}

\textbf{Keywords:} Human--AI interaction; Large Language Models; LLMography; AI provenance; AI auditability; education; prompt analysis; academic integrity; traceability; reproducibility; Human--AI collaboration.

\newpage
\section{Introduction}

Large Language Models are increasingly embedded in educational, professional, and technical workflows. Students use them to draft reports, generate diagrams, debug code, refine arguments, translate content, design architectures, and prepare technical documentation. Developers use AI assistants to explore repositories, generate functions, fix errors, write tests, and produce documentation. Researchers and professionals use them to summarize literature, structure ideas, write documents, and support decision-making.

This transformation creates a major challenge for education, research, software engineering, and organizational governance. The final artifact alone no longer tells the full story. A project report, a LaTeX document, a UML diagram, or a code module may result from a complex sequence of Human--AI interactions involving prompts, responses, corrections, revisions, source checks, tool executions, and final validation.

Most current discussions about AI-assisted work focus on a narrow question: \textit{Was AI used?} This question is important but incomplete. In many contexts, a more meaningful question is: \textit{How was AI used?} Did the human provide clear direction? Did the user correct the AI? Was the output accepted passively? Were sources validated? Was the final artifact traceable to the conversation history? Can the production process be audited?

This paper proposes \textbf{LLMography} as a framework for answering these questions. LLMography transforms Human--AI conversations into structured indicators of traceability, human contribution, AI dependency, reproducibility, and auditability.

The central principle is:

\begin{quote}
\textit{In the AI age, outputs are not enough. We need the history of interaction.}
\end{quote}

LLMography does not aim to accuse users, detect misconduct, or judge whether AI use is morally acceptable. Instead, it provides evidence-based indicators derived from the available conversation trace. It is intended to help teachers, researchers, students, developers, organizations, and auditors understand the process through which an AI-assisted artifact was produced.

The guiding statement of the framework is:

\begin{quote}
\textit{LLMography does not judge whether AI was used. It explains how AI was used.}
\end{quote}

The term \textbf{LLMography}, the analogy with bibliography and webography, the slogan \textit{``In the AI age, outputs are not enough. We need the history of interaction,''} and the overall research direction were initiated and defined by the human author. AI assistance was used as a support tool for drafting, structuring, prototyping, LaTeX preparation, and refinement.

This paper makes the following contributions:

\begin{enumerate}
    \item It introduces \textbf{LLMography} as a conceptual term and computational framework for documenting Human--AI interaction histories.
    \item It positions LLMography by analogy with bibliography and webography, while extending the idea toward dynamic Human--AI co-construction.
    \item It defines a set of KPIs for measuring prompt quality, human direction, AI dependency, traceability, auditability, and privacy risk.
    \item It presents a prototype system that generates structured LLMography KPI reports from conversation traces.
    \item It reports a preliminary exploratory evaluation using 19 anonymized student audit reports.
    \item It applies LLMography to the writing process of the present paper itself.
    \item It discusses educational, ethical, methodological, and technical implications of moving from AI detection to AI-use documentation.
\end{enumerate}

\section{Conceptual Positioning: From Bibliography and Webography to LLMography}

A review of current academic and professional discourse suggests that the documentation of Human--AI interaction histories is still described through fragmented and mostly technical terms. Existing expressions include \textit{prompt logging}, \textit{prompt tracking}, \textit{AI prompt documentation}, \textit{conversational history}, \textit{AI-assisted workflow documentation}, and, in educational contexts, appendices or declarations of AI prompts. These terms are useful, but they generally remain descriptive, procedural, or administrative.

For example, institutional guidance increasingly recommends that users acknowledge or cite generative AI tools. Some academic style guides explain how generative AI outputs should be referenced, and some educational institutions ask students to declare AI use or provide prompt appendices \cite{apa2023chatgpt,apa2025genai}. These practices show the emergence of a transparency need. However, they remain largely centered on citation, disclosure, or prompt recording rather than on the full interaction history.

LLMography differs from these fragmented approaches by proposing a unified concept. The term is intentionally constructed by analogy with \textbf{bibliography} and \textbf{webography}: a bibliography lists the books and articles that informed a work; a webography lists the web resources consulted during knowledge production; and a LLMography documents the dynamic trajectory of interaction between a human and a Large Language Model.

The suffix \textit{-graphy}, derived from the Greek \textit{graphein}, meaning ``to write'' or ``to trace,'' is central to the concept. LLMography is not merely a list of prompts. It is a structured trace, or map, of the Human--AI co-construction process. It records prompts, AI responses, human corrections, accepted or rejected suggestions, contextual sources, tool calls, revisions, validation steps, and final outputs.

This distinction is important. A prompt appendix may show what was asked. A prompt log may show that an interaction occurred. But LLMography aims to show how the final artifact emerged from an interaction process. It therefore shifts the focus from isolated prompts to interaction provenance, from AI detection to AI-use documentation, and from output-centered evaluation to process-centered auditability.

To the best of our knowledge, the literature has not yet consolidated these practices under a single pedagogically oriented concept comparable to bibliography or webography. LLMography is proposed as such a concept: a provenance layer for trustworthy Human--AI collaboration.

\section{Related Work}

LLMography is positioned at the intersection of provenance modeling, Human--AI interaction, educational assessment, academic integrity, and AI-use documentation.

\subsection{Provenance and Traceability}

Provenance research provides a formal foundation for describing how artifacts are produced. The W3C PROV model defines core concepts such as entities, activities, and agents involved in the production and transformation of artifacts \cite{w3c_prov_dm_2013,w3c_prov_overview_2013}. This is highly relevant to AI-assisted work because Human--AI collaboration can be understood as a sequence of activities involving human agents, AI agents, contextual sources, tools, and final outputs.

However, general provenance models do not directly specify how to represent the specific dynamics of Human--LLM interaction: prompts, responses, corrections, accepted suggestions, rejected alternatives, prompt evolution, and human validation. LLMography can therefore be seen as a domain-specific extension of provenance thinking for Human--AI interaction \cite{missier2013w3cprov}.

\subsection{Human--AI Interaction}

Research in Human--AI interaction emphasizes transparency, controllability, feedback, user understanding, and appropriate reliance. Guidelines for Human--AI interaction have highlighted the importance of helping users understand AI behavior, recover from errors, and maintain appropriate control over AI systems \cite{amershi2019guidelines}.

LLMography builds on this perspective by focusing not only on interface design, but on the post-hoc documentation and audit of the Human--AI collaboration process. It asks whether a conversation trace can reveal how the user controlled the interaction, how the AI contributed, and whether the final artifact can be reconstructed.

\subsection{Generative AI in Higher Education}

Generative AI has created major challenges in higher education. Students can now use LLMs to produce essays, code, reports, presentations, diagrams, and analyses. This challenges traditional assessment methods and academic integrity policies. Recent work suggests that banning or detecting AI use is insufficient, and that institutions need more nuanced frameworks for integrating AI into assessment and learning \cite{ardito2023contra,wang2023generative,ogunleye2024higher,furze2024aias}.

LLMography contributes to this discussion by offering a process-centered approach. Instead of asking only whether a student used AI, LLMography asks how the student used AI and whether the interaction demonstrates human direction, critical thinking, source validation, and traceability.

\subsection{Academic Integrity and AI Detection}

AI detection tools have been criticized for reliability issues, ethical risks, false positives, and limited pedagogical value. A detection-based approach often reduces the problem to a binary classification: AI-generated or human-written. This is increasingly problematic because many real-world outputs are neither fully human nor fully AI-generated. They are produced through iterative collaboration.

LLMography proposes a complementary alternative: documenting the Human--AI process. This does not replace academic integrity policies, but it can enrich them with evidence about interaction quality and human oversight \cite{tan2024shaping,kofinas2025impact,bittle2025generative}.

\section{The LLMography Framework}

LLMography is a framework for documenting, analyzing, and auditing Human--AI conversations. It treats the interaction history as a first-class artifact.

A LLMography analysis takes as input a Human--AI conversation trace, such as a shared chat, exported transcript, copied interaction, or platform log. It produces a structured report containing a conversation overview, interaction summary, KPI dashboard, Human--AI contribution analysis, strengths, weaknesses and risks, audit interpretation, limitations, and a recommended LLMography label.

\subsection{Core Principle}

The core principle of LLMography is that AI-assisted artifacts should not be evaluated only through their final form. The process matters. The interaction history can reveal whether the human controlled the process, whether the AI was used passively or critically, whether the output was iteratively refined, and whether the final artifact is traceable.

\subsection{LLMography Labels}

\begin{table}[h]
\centering
\caption{LLMography labels.}
\begin{tabular}{ll}
\toprule
Label & Description \\
\midrule
A & Minimal AI assistance \\
B & AI-assisted, human-directed \\
C & Human--AI co-produced \\
D & AI-dominant production \\
E & Insufficient traceability \\
\bottomrule
\end{tabular}
\end{table}

These labels are not moral judgments. They summarize the visible collaboration pattern.

\subsection{KPI Categories and Scoring}

LLMography estimates conversation structure KPIs, prompting quality KPIs, human contribution KPIs, AI dependency KPIs, traceability and audit KPIs, and privacy and risk KPIs. Scores are interpreted on a 0--100 scale, where 0--20 is very weak, 21--40 is weak, 41--60 is moderate, 61--80 is good, and 81--100 is strong. KPI values are evidence-based estimates derived from the available trace, not absolute proofs of authorship or correctness.

\section{Methodology}

\subsection{Prototype System}

A lightweight prototype called \textbf{LLMography.ai} has been implemented as an AI-based web analyzer. The initial MVP allows a user to paste a Human--AI conversation transcript or shared conversation link and generate a KPI report. The prototype follows a simple pipeline: conversation input, conversation parsing, LLMography analysis, KPI extraction, and dashboard display. The MVP was implemented using a lightweight FastAPI backend, HTML/CSS frontend, Docker deployment, and an OpenAI model for analysis.

\begin{figure}[h]
\centering
\includegraphics[width=0.95\linewidth]{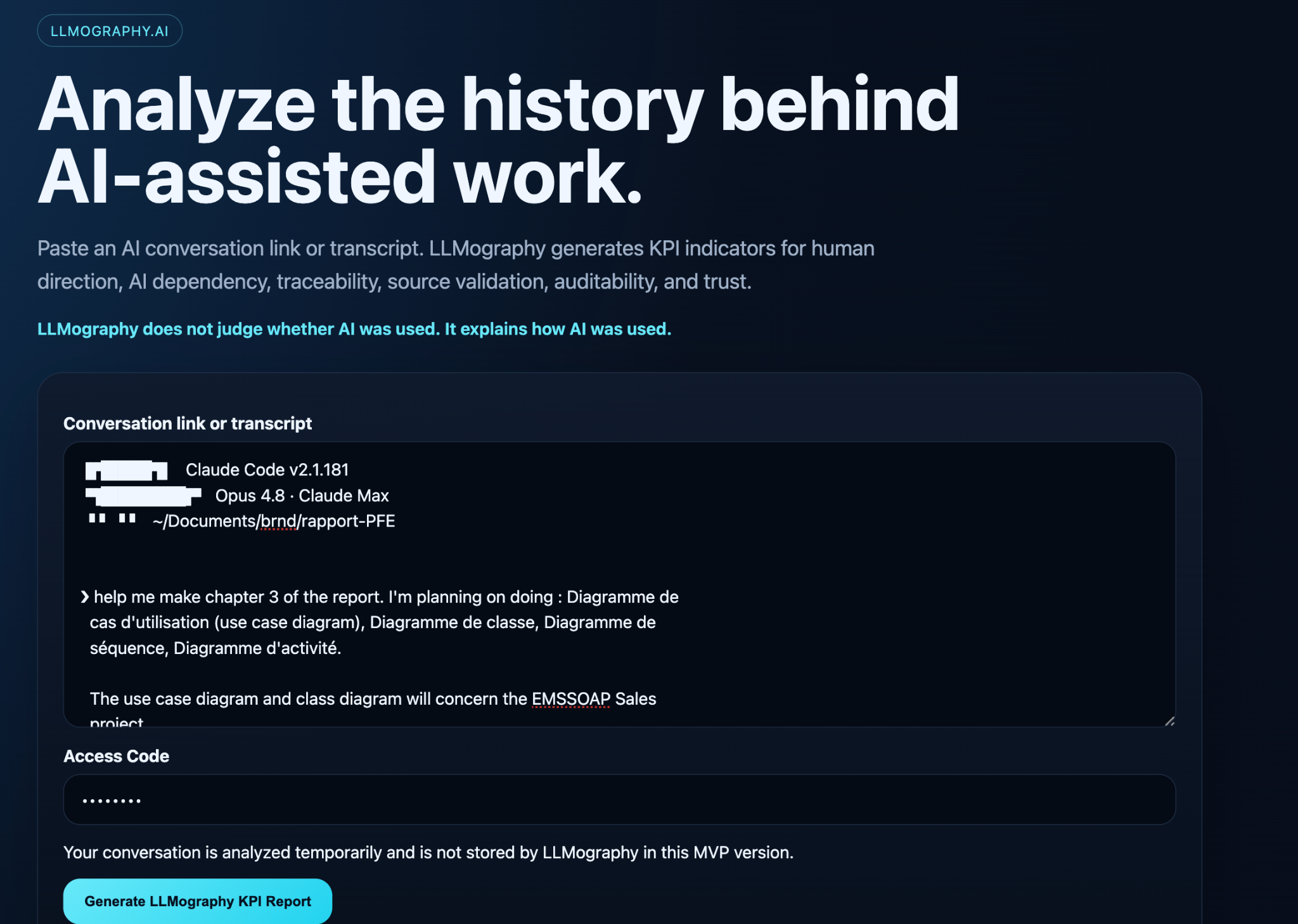}
\caption{LLMography MVP input interface for submitting a Human--AI conversation link or transcript.}
\label{fig:mvp-input}
\end{figure}

\begin{figure}[h]
\centering
\includegraphics[width=0.95\linewidth]{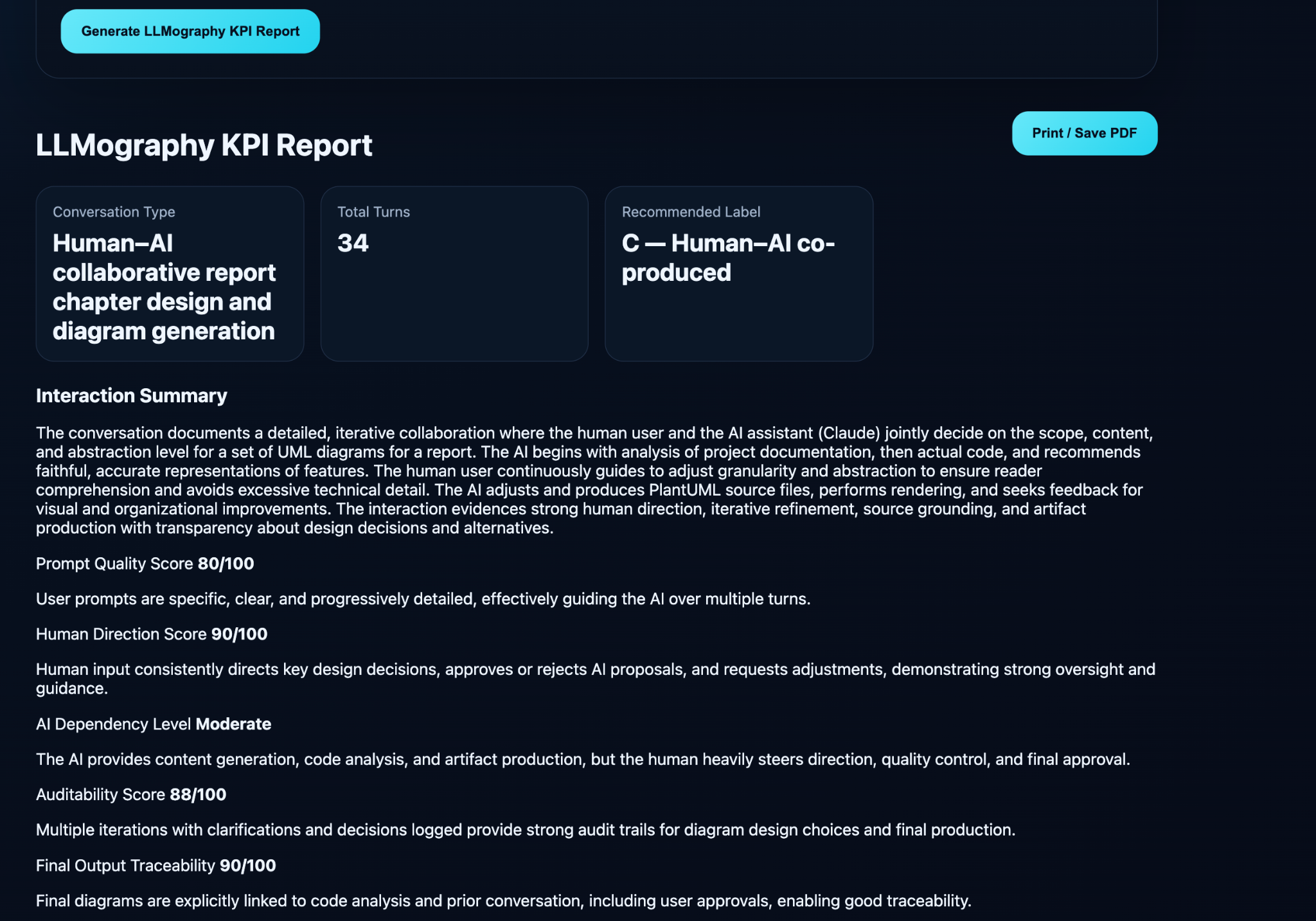}
\caption{Example of a LLMography KPI report generated by the MVP, including conversation type, total turns, label, interaction summary, and core KPI scores.}
\label{fig:mvp-report-summary}
\end{figure}

\begin{figure}[h]
\centering
\includegraphics[width=0.95\linewidth]{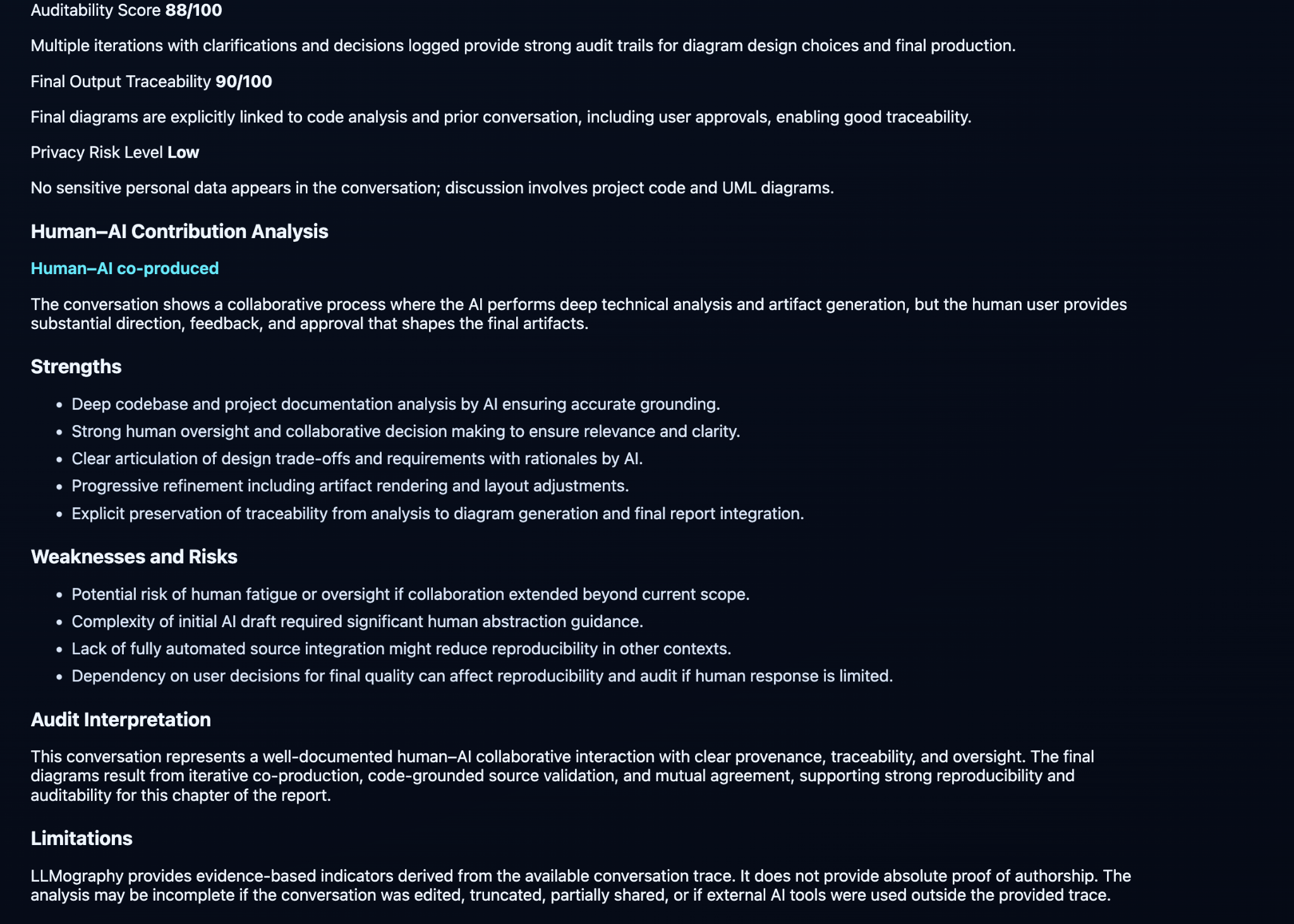}
\caption{Example of the detailed LLMography audit interpretation, including human--AI contribution analysis, strengths, weaknesses, risks, and limitations.}
\label{fig:mvp-detailed-audit}
\end{figure}

Figure~\ref{fig:mvp-input} shows the MVP input interface, where a user pastes an AI conversation link or transcript. Figures~\ref{fig:mvp-report-summary} and~\ref{fig:mvp-detailed-audit} show examples of the generated KPI report and detailed audit interpretation.

The LLMography Analyzer system prompt instructs the model to parse the conversation into user and AI turns, identify the task type, build an interaction map, compute or estimate KPI values, explain each KPI in plain language, generate an audit report, state uncertainty and limitations, and avoid claiming absolute proof of authorship. The MVP is designed with minimal data retention: conversations are analyzed temporarily and are not stored by default.
\subsection{Study Context}

We conducted a preliminary exploratory evaluation of LLMography using Human--AI conversation audit reports provided by 19 engineering students. The participants were final-year engineering students engaged in their end-of-study project phase, working on technical, academic, and software-related tasks.

The students voluntarily agreed to participate in the experiment and accepted the principle of auditing their AI-assisted conversations. An oral discussion was conducted with them to explain the objective of the experiment and to confirm that the goal was not punitive, but pedagogical and scientific.

The objective was to evaluate whether real Human--AI conversation traces, produced during the final-year project phase, could be transformed into measurable indicators of collaboration quality, human direction, AI dependency, traceability, and auditability.

\subsection{Dataset}

The current study is based on 19 available LLMography KPI audit reports. The conversations cover project report generation, LaTeX editing and compilation, UML and PlantUML diagram generation, BPMN architecture modeling, software development and debugging, backend authentication and RBAC design, data science reporting, technical documentation, academic chapter drafting, and UI prototyping.

\subsection{Consent and Anonymization}

Students were informed orally about the purpose of the experiment and agreed to participate. For future publication, written consent and anonymization procedures should be formally completed. In the reported analysis, the focus is on interaction patterns and aggregate KPI indicators rather than individual identity. Student names, email addresses, phone numbers, company-sensitive identifiers, credentials, and private project data should be removed or anonymized before any public release.

A recommended ethical statement is:

\begin{quote}
\textit{This study does not aim to detect misconduct. It aims to document and understand how students use AI in real academic workflows.}
\end{quote}

\newpage
\subsection{KPI Extraction}

Each conversation was analyzed using the LLMography prototype, which generated a structured KPI report. The present paper aggregates Total Turns, Prompt Quality Score, Human Direction Score, AI Dependency Level, Auditability Score, Final Output Traceability, Privacy Risk Level, and Recommended LLMography Label.

\section{Results and Discussion}

\subsection{Aggregate Conversation Statistics}

The 19 available audit reports contain a total of 462 conversation turns.

\begin{table}[h]
\centering
\caption{Aggregate conversation statistics.}
\begin{tabular}{lr}
\toprule
Metric & Value \\
\midrule
Number of analyzed audit reports & 19 \\
Total conversation turns & 462 \\
Average turns per conversation & 24.3 \\
Median turns per conversation & 22 \\
Minimum turns & 1 \\
Maximum turns & 106 \\
\bottomrule
\end{tabular}
\end{table}

\subsection{Aggregate KPI Scores}

\begin{figure}[h]
\centering
\includegraphics[width=0.85\linewidth]{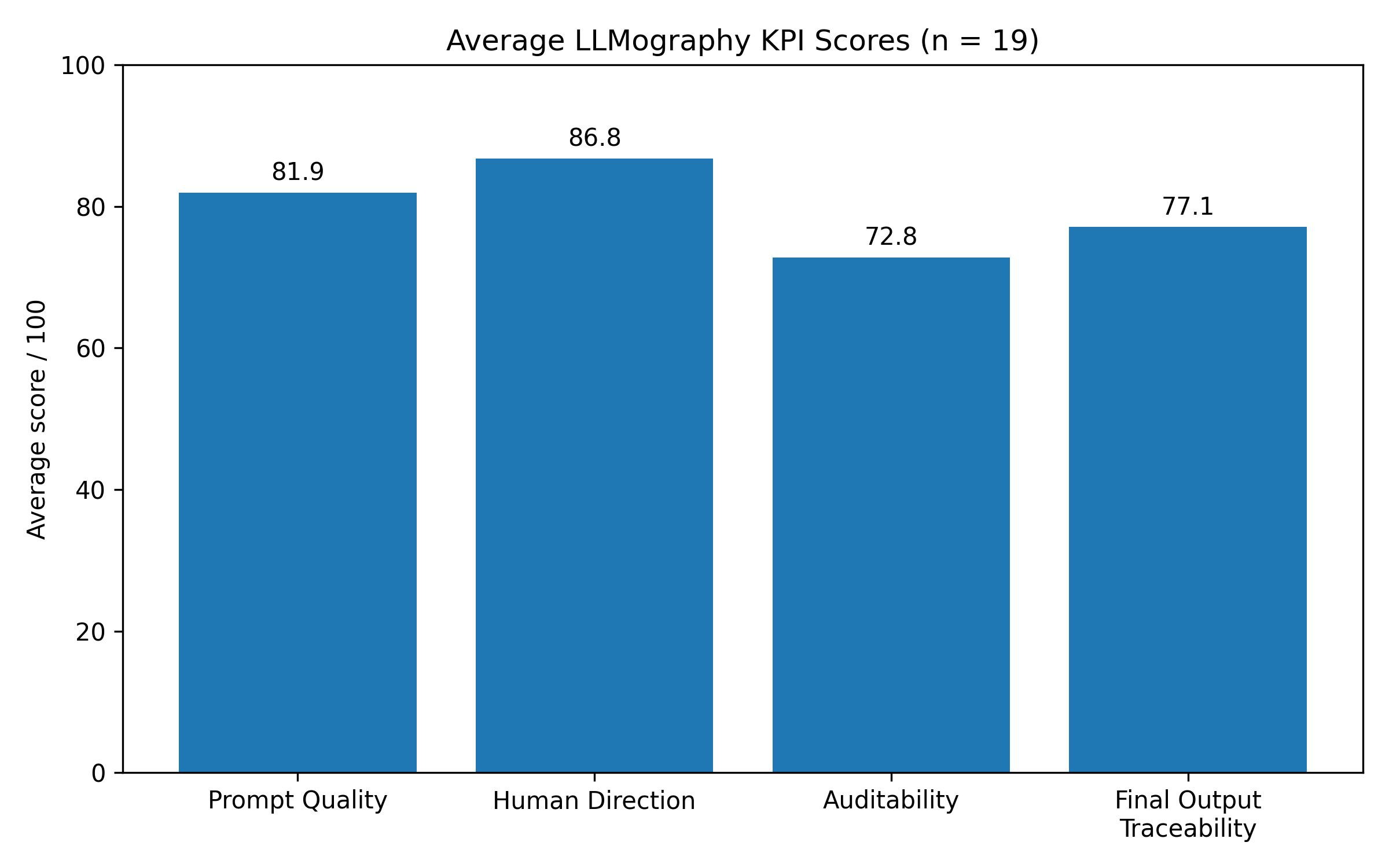}
\caption{Average LLMography KPI scores across the 19 analyzed audit reports.}
\label{fig:kpi-means}
\end{figure}

The average scores across the 19 available reports are shown in Table~\ref{tab:kpi-averages} and Figure~\ref{fig:kpi-means}.

\begin{table}[h]
\centering
\caption{Average KPI scores.}
\label{tab:kpi-averages}
\begin{tabular}{lr}
\toprule
KPI & Average Score \\
\midrule
Prompt Quality Score & 81.9/100 \\
Human Direction Score & 86.8/100 \\
Auditability Score & 72.8/100 \\
Final Output Traceability & 77.1/100 \\
\bottomrule
\end{tabular}
\end{table}

These results suggest that students were generally capable of providing clear prompts and maintaining strong direction over the AI-assisted process. The Human Direction Score is particularly high, indicating that many interactions were not passive uses of AI but rather guided collaborations.

\subsection{Distribution of LLMography Labels}

\begin{figure}[h]
\centering
\includegraphics[width=0.85\linewidth]{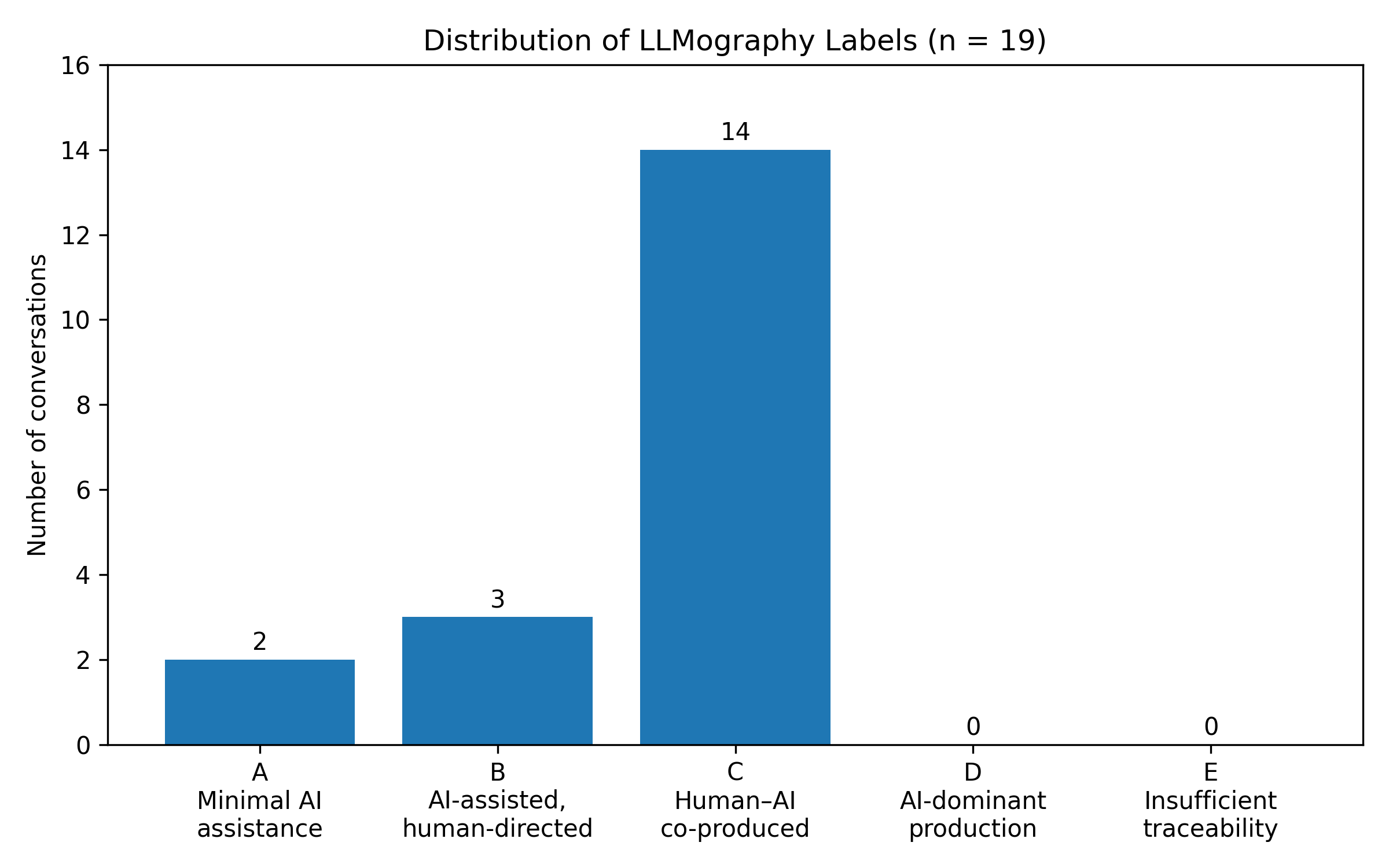}
\caption{Distribution of LLMography labels across the 19 analyzed audit reports.}
\label{fig:label-distribution}
\end{figure}

The label distribution is shown in Table~\ref{tab:labels} and Figure~\ref{fig:label-distribution}.

\begin{table}[h]
\centering
\caption{Distribution of LLMography labels.}
\label{tab:labels}
\begin{tabular}{llr}
\toprule
Label & Meaning & Count \\
\midrule
A & Minimal AI assistance & 2 \\
B & AI-assisted, human-directed & 3 \\
C & Human--AI co-produced & 14 \\
D & AI-dominant production & 0 \\
E & Insufficient traceability & 0 \\
\bottomrule
\end{tabular}
\end{table}

Most conversations were classified as \textbf{Label C --- Human--AI co-produced}. This indicates that the majority of student interactions involved iterative collaboration where the human provided direction, feedback, constraints, or validation while the AI contributed drafting, coding, formatting, diagramming, or technical execution.

\subsection{AI Dependency Levels and Privacy Risk}

Most conversations show moderate AI dependency: 2 low, 15 moderate, 2 high, and 0 very high. This means that the AI played a significant role in content generation, coding, editing, or documentation, but the human user retained visible control through prompts, corrections, objectives, and validation.

All available reports classified privacy risk as low. The conversations mainly involved project technical data, report writing, LaTeX formatting, diagram generation, and software development tasks. No major personal or sensitive data exposure was visible in the analyzed reports. This finding is preliminary and depends on the anonymized and available traces.

\subsection{Qualitative Findings}

Many students provided specific instructions, corrections, formatting requirements, project context, and progressive refinements. This led to high Human Direction Scores. In conversations involving UML diagrams, BPMN diagrams, LaTeX reports, or software debugging, students often guided the AI toward specific expected artifacts. This suggests that AI use in student workflows cannot be reduced to passive generation.

The AI often performed tasks such as generating LaTeX code, compiling or debugging reports, generating PlantUML or BPMN artifacts, editing project documentation, analyzing backend architecture, proposing code fixes, creating report sections, integrating images and figures, and generating structured explanations. This indicates that students used AI not only as a writing assistant, but also as a technical execution partner.

Conversations with command logs, file inspection traces, compilation checks, explicit artifact names, and progressive edits generally had higher auditability and final output traceability. Conversely, single-turn conversations were less useful for studying iterative collaboration, even when the prompt itself was detailed.

Several audit reports mention limited external source validation. AI-assisted work may be well-structured and traceable while still lacking strong evidence validation. Future LLMography versions should therefore distinguish between interaction traceability, source validation, factual reliability, and artifact reproducibility.

\subsection{Discussion}

The preliminary results support the central argument of LLMography: AI transparency should not be limited to detecting whether AI-generated text exists. In educational contexts, the more important question may be how the student interacted with the AI. A student who provides strong direction, corrects the AI, integrates project-specific knowledge, verifies outputs, and iteratively refines an artifact is engaging in a different process from a student who submits a one-shot AI-generated answer.

LLMography can help educators move from prohibition-based approaches to evidence-based AI literacy. Instead of asking students only to declare AI use, educators may ask for an interaction trace or LLMography report showing the role of AI, the role of the student, the quality of prompting, the level of critical thinking, the traceability of the final artifact, and the presence or absence of validation.

The study also shows that conversation traces can support auditability when they include sufficient detail. Interactions involving file operations, command outputs, code changes, compilation logs, or explicit artifact generation are more auditable than vague or single-turn interactions. This suggests that future AI-assisted systems should encourage structured logging, artifact linkage, source tracking, and versioning.

\subsection{Threats to Validity}

The current evaluation is based on 19 available audit reports. This is a small exploratory sample and should not be generalized without further studies. The conversations were submitted by students who agreed to participate, which may introduce selection bias. The analysis also relies on available conversation traces; external tools, deleted messages, unshared prompts, or offline edits may not be visible.

Some KPI scores are estimated from qualitative evidence in the conversation. Future work should improve score calibration, inter-rater validation, and reproducibility of KPI computation. The present study does not compare LLMography scores against an independent human-coded ground truth. Future studies should involve multiple reviewers and compare automated KPI extraction with expert annotations.

\subsection{Ethical Considerations}

LLMography must be designed and used carefully. Conversation histories may contain sensitive information, private reasoning, project details, or personal data. Therefore, ethical use requires informed consent, anonymization, data minimization, clear purpose limitation, secure storage, right to withdraw before final anonymization, and non-punitive use in educational contexts.

The purpose of this study is not to accuse students of improper AI use. It is to better understand Human--AI collaboration in real academic and technical workflows.

In future versions, LLMography should include privacy-by-design mechanisms such as automatic redaction, local-first analysis, hashed evidence, selective disclosure, access-controlled reports, and separate public and private audit modes.

\section{Application of LLMography to the Writing of This Paper}

In addition to the student audit reports, the preparation of this manuscript itself provides a real-world application of the LLMography approach. The author used a Human--AI conversation with ChatGPT to progressively develop the concept, structure the framework, refine the terminology, generate early drafts, organize the KPI model, prepare the LaTeX manuscript, and discuss ethical and publication-related issues.

This interaction was not treated as hidden assistance, but as a traceable Human--AI collaboration. The conversation history shows that the human author initiated the concept of LLMography, chose the term \textbf{LLMography}, formulated the central slogan --- \textit{``In the AI age, outputs are not enough. We need the history of interaction.''} --- defined the analogy with bibliography and webography, provided the educational and experimental context, selected the research direction, supplied the student audit reports, validated the KPI interpretation, and retained final responsibility for the manuscript.

The AI assistant contributed by supporting structuring, reformulation, code generation, LaTeX preparation, figure generation, and methodological refinement. These contributions were assistive and instrumental. They did not originate the concept, choose the name, define the slogan, determine the research agenda, design the experimental context, or assume responsibility for the scientific claims.

Applying LLMography to this author--AI conversation led to the qualitative classification shown in Table~\ref{tab:self-application}.

\begin{table}[h]
\centering
\caption{LLMographic self-application to the writing of this paper.}
\label{tab:self-application}
\begin{tabular}{ll}
\toprule
Indicator & Value \\
\midrule
Conversation Type & Research ideation, prototype development, academic writing \\
Recommended Label & B --- AI-assisted, human-directed \\
Precise characterization & Human-originated, human-directed, AI-assisted  \\
Prompt Quality Score & 88/100 \\
Human Direction Score & 94/100 \\
Human Initiative Score & 96/100 \\
AI Dependency Level & Moderate \\
Human Oversight Score & 90/100 \\
Auditability Score & 90/100 \\
Final Output Traceability & 92/100 \\
Privacy Risk Level & Moderate--Low \\
\bottomrule
\end{tabular}
\end{table}

This self-application illustrates the central claim of the paper: the value of AI transparency does not lie only in declaring that AI was used, but in documenting how it was used. The manuscript was therefore not only written with AI assistance; its own writing process became an object of LLMographic analysis.

This case also demonstrates the difference between AI-assisted writing and AI-dominant authorship. The human author provided the original concept, the name LLMography, the slogan, the research motivation, the experimental setting, the student audit data, the methodological orientation, the validation, and the scientific responsibility. The AI assistant accelerated drafting, structuring, technical preparation, LaTeX generation, figure preparation, and formulation. The resulting process is best described as \textbf{human-originated, human-directed, AI-assisted }.

\section{Conclusion and Future Work}

This paper introduced LLMography, a framework for transforming Human--AI conversations into measurable indicators of traceability, human contribution, AI dependency, reproducibility, and auditability. By analogy with bibliography and webography, LLMography proposes that the interaction history between humans and LLMs should become a first-class artifact.

A preliminary evaluation using 19 available student-provided audit reports suggests that many AI-assisted academic and technical workflows are not simply AI-generated outputs, but iterative Human--AI co-productions. The results show high average Human Direction Score, strong Prompt Quality Score, moderate AI dependency in most cases, and generally good traceability.

The self-application of LLMography to the writing of this paper further illustrates the practical relevance of the framework. It shows that AI-assisted writing can be made transparent when the history of interaction is documented, analyzed, and interpreted. In this case, the process was human-originated, human-directed, and AI-assisted.

These findings support the central thesis of LLMography: the future of AI transparency should not focus only on final outputs, but on the history of interaction that produced them.

\begin{quote}
\textit{In the AI age, outputs are not enough. We need the history of interaction.}
\end{quote}

Future work will focus on seven directions. First, the prototype will be extended to support structured export of KPI results in CSV and JSON formats. Second, a larger dataset of student conversations will be collected with formal written consent and systematic anonymization. Third, a dashboard will be developed to visualize aggregate distributions of AI dependency, human direction, prompt quality, and auditability. Fourth, the scoring system will be validated through comparison with human expert annotations. Fifth, LLMography will be extended to support integration with learning management systems, Git repositories, document editors, and AI coding assistants. Sixth, the system will incorporate source validation and citation consistency scoring to distinguish interaction traceability from factual reliability. Seventh, future work will explore standardization of LLMography records as interoperable metadata for AI-assisted work.

\section*{Acknowledgments}

The author would like to thank Doha Bousmah, a Computer Science student at Illinois Institute of Technology (IIT), for her careful reading of the manuscript and for providing valuable feedback on the clarity, positioning, and presentation of the LLMography concept, as well as on the discussion of Human--AI interaction in academic work.

The author also gratefully acknowledges the 19 engineering students who voluntarily participated in the preliminary LLMography experiment during their final-year project phase. Their participation and consent made it possible to conduct the initial exploratory analysis presented in this paper.

\section*{Declaration of AI-Assisted Writing and Human Intellectual Contribution}

The author used ChatGPT as a writing and research-assistance tool during the preparation of this manuscript. The tool was used to help structure the article, reformulate sections, generate LaTeX drafts, prepare figures, and improve clarity. However, the intellectual origin and strategic direction of the work remain entirely human-led.

The human author initiated and defined the core concept of \textbf{LLMography}. The author chose the term \textbf{LLMography}, established its analogy with bibliography and webography, formulated the central slogan --- \textit{``In the AI age, outputs are not enough. We need the history of interaction.''} --- and defined the general research direction, pedagogical motivation, experimental context, KPI-based approach, and audit methodology.

The author also designed the initial experimental setting, collected and provided the student audit reports, interpreted the educational relevance of the results, selected the arXiv orientation, and retained full responsibility for the scientific claims, methodological choices, ethical framing, and final manuscript.

The AI assistant contributed to the execution and acceleration of the work by supporting structuring, reformulation, drafting, LaTeX preparation, figure generation, code generation, and methodological refinement. These contributions were assistive and instrumental; they did not originate the concept, choose the name, define the slogan, determine the research agenda, or assume responsibility for the work.

All AI-assisted text was reviewed, edited, and validated by the author. The AI tool is not listed as an author because it cannot take responsibility for the content, approve the final manuscript, guarantee the integrity of the research, or be accountable for the work.

Consistent with the LLMography framework proposed in this paper, the Human--AI conversation that supported the preparation of this manuscript was itself analyzed as a real application of LLMography. The interaction was classified as \textbf{Label B --- AI-assisted, human-directed }, with strong human direction, moderate AI dependency, high auditability, and high final output traceability. More precisely, the process is best described as \textbf{human-originated, human-directed, AI-assisted}.

\appendix
\newpage
\section{Preliminary KPI Summary}

\begin{longtable}{p{0.08\linewidth}p{0.31\linewidth}rcccrr}
\caption{Preliminary KPI summary for the 19 analyzed audit reports.}\\
\toprule
ID & Task Type & Turns & Label & Prompt & Human & Audit & Trace \\
 & & & & Quality & Direction & ability & ability \\
\midrule
\endfirsthead
\toprule
ID & Task Type & Turns & Label & Prompt & Human & Audit & Trace \\
 & & & & Quality & Direction & ability & ability \\
\midrule
\endhead
C01 & Project report and database seeding & 26 & C & 75 & 80 & 80 & 80 \\
C02 & UML diagram and report chapter design & 34 & C & 80 & 90 & 88 & 90 \\
C03 & Report generation & 10 & C & 75 & 80 & 60 & 65 \\
C04 & KPI extraction software planning & 10 & C & 90 & 95 & 75 & 80 \\
C05 & LaTeX data science report & 106 & C & 90 & 88 & 72 & 80 \\
C06 & Software development & 40 & C & 80 & 85 & 75 & 80 \\
C07 & Backend debugging specification & 1 & B & 95 & 95 & 90 & 90 \\
C08 & LaTeX editing and compilation & 22 & C & 80 & 85 & 75 & 80 \\
C09 & Procedural instruction guide & 1 & A & 85 & 100 & 10 & 0 \\
C10 & Technical design specification & 1 & A & 90 & 95 & 90 & 100 \\
C11 & PlantUML and bibliography & 12 & C & 75 & 85 & 75 & 80 \\
C12 & SpeakSmart design and prototyping & 22 & C & 90 & 88 & 78 & 80 \\
C13 & Technical project overview & 2 & B & 80 & 85 & 65 & 75 \\
C14 & Document enhancement and compilation & 46 & C & 70 & 85 & 70 & 75 \\
C15 & Academic report drafting & 6 & B & 70 & 75 & 70 & 80 \\
C16 & Document restructuring & 57 & C & 75 & 80 & 60 & 65 \\
C17 & LaTeX formatting assistance & 36 & C & 70 & 85 & 85 & 90 \\
C18 & BPMN and architecture documentation & 28 & C & 92 & 88 & 80 & 85 \\
C19 & Authentication and RBAC implementation & 2 & C & 95 & 85 & 85 & 90 \\
\bottomrule
\end{longtable}

\bibliographystyle{plain}
\bibliography{references}

\end{document}